\documentclass[twocolumn]{aastex63}

\usepackage{fdsymbol}
\usepackage{color}
\usepackage{float}

\usepackage{graphicx}	
\usepackage{amsmath}	
\usepackage{amssymb}	

\received{2019 September 11}
\revised {2019 October 30}
\accepted{2019 November 8}
\published{2019 November 27}

\submitjournal{The Astrophysical Journal Letters}

\shorttitle{Laboratory Experiments on the Motion of Dense Dust Clouds}
\shortauthors{Schneider and Wurm}

\begin{document}

\title{\Large{Laboratory Experiments on the Motion of Dense Dust Clouds in Protoplanetary Disks}}

\correspondingauthor{Niclas Schneider}
\email{niclas.schneider@uni-due.de}

\author{Niclas Schneider}
\affiliation{Faculty of Physics \\
University of Duisburg-Essen \\
Lotharstr. 1, 47057 Duisburg, Germany}

\author{Gerhard Wurm}
\affiliation{Faculty of Physics \\
University of Duisburg-Essen \\
Lotharstr. 1, 47057 Duisburg, Germany}



\begin{abstract}

{In laboratory experiments, we study the motion of levitated, sedimenting clouds of sub-mm grains at low ambient pressure and at high {solid-}to-gas ratios $\epsilon$. The experiments show a collective behavior of particles, i.e. grains in clouds settle faster than an isolated grain. In collective particle clouds, the sedimentation velocity linearly depends on $\epsilon$ and linearly depends on the particle closeness $C$. However, collective {behavior} only sets in at a critical value $\epsilon_{\rm crit}$ which linearly increases with the experiment Stokes number St. For St $<$ 0.003 particles always behave collectively. For large Stokes numbers, large solid-to-gas ratios are needed to trigger collective {behavior}, e.g. $\epsilon_{\rm crit} = 0.04$ at St = 0.01.  Applied to protoplanetary disks, {particles in dense environments will settle faster. In balance with upward gas motions (turbulent diffusion, convection) the thickness of the midplane particle layer will be smaller than calculated based on individual grains, especially for dust. For pebbles, large solid-to-gas ratios are needed to trigger instabilities based on back-reaction.}}

\end{abstract}





\section{Introduction}

{Planet formation starts with sticking collisions of dust in protoplanetary disks \citep{blumwurm2008}. Relative velocities between the solids are provided by sedimentation to the midplane, radial and transversal drifts, and turbulence \citep{Birnstiel2016}. Collisional grain growth might proceed at least to millimeter grain size before bouncing dominates the outcome of a collision \citep{zsom2010, Demirci2017}.}

{It is the interaction or coupling between the solid grains and the gas that sets the collision velocities. Beyond collisions, gas-grain coupling also determines the dust scale height or how far particles sediment to the midplane in balance with turbulent mixing \citep{Birnstiel2016, Pignatale2017}. Gas-grain coupling is also important for the radial inward drift, especially of decimeter- to meter-sized bodies \citep{Weidenschilling1977}, and it is a major part in trapping particles in pressure bumps \citep{whipple1972}.} 
While for some aspects particles can be considered as tracer particles -- individual grains with no influence on the gas motion itself -- dense particle clouds require a more complex treatment. 

{In recent years, particle-gas feedback was suggested to promote particle concentration that eventually lead to gravitational collapse into planetesimals \citep{Youdin2005, JohansenYoudin2007, gonzalez2017, Dipierro2018, Squire2018SI}. Concentration mechanisms depend strongly on the particles' Stokes numbers, the metallicity, and the solid-to-gas ratio (e.g. \cite{bai2010c, carrera2015, yang2017, Squire2018SI}). In any case, these mechanisms might take over from collisional growth at pebble size to form planetesimals.}
 
Assisting this numerical work relying on particle-gas feedback mechanisms, we investigate the motion of dense particle clouds in a thin gas in laboratory experiments here.

A basic concept in a simple system of one particle in an unlimited reservoir of gas is that the grain needs a certain gas-grain friction time $\tau_f$ to follow any change in gas motion or react to any external force and reach equilibrium between external force and friction. 
The flow around the particle can be divided in molecular flow ($\rm Kn \gg 1$) {determined by} Epstein drag and continuum flow ($\rm Kn \ll 1$) {determined by} Stokes drag where Kn is the Knudsen number with the mean free path $\lambda$ and particle radius $r$, 
\begin{equation}
\mathrm{Kn} = \frac{\lambda}{r}.
\label{Knudsen}
\end{equation}
The stationary sedimentation speed of a grain is given by 
\begin{equation}
v_0 = \tau_f \cdot g, 
\label{tau_f}
\end{equation}
where $g$ is the gravitational acceleration, i.e. the vertical component of a star's gravity in a protoplanetary disk.

The motion of an individual particle in a cloud of many particles can only be treated this way to a certain limit. It requires that {back-reaction} on the gas and feedback from this {back-reaction} to the other grains can be neglected. This is only true for an isolated particle, i.e. for low  {solid-}to-gas mass ratios and low volume filling factors. In protoplanetary disks, the canonical  {solid-}to-gas ratio of 0.01 can change by sedimentation and other concentration mechanisms by several orders of magnitude, while the volume filling factor remains low ($< 10^{-6}$) \citep{Klahr2018}.

Grains in a dense cloud might just effectively behave like a larger particle, moving faster like the individual grains \citep{JohansenYoudin2007}. {Eventually}, collective behavior might lead to planetesimal formation \citep{Johansen2007, Chiang2010, Klahr2018}.

In \cite{schneider2019}, we studied the transition from test particle to collective behavior in {a} levitation experiment, analyzing the {free-fall} velocity of grains in a cloud. 
We {empirically} found that the sedimentation velocity depends on what we call sensitivity factor $F_S$ and the closeness $C$ of the individual particle as
\begin{equation}
    v_s = v_0 + F_S \cdot C.
    \label{sedvel}
\end{equation}
{The closeness $C$ of a particle is constructed from the interparticle distances $r_j-r$ between the grain and all other particles $j$ as}
\begin{equation}
    C = \sum_{j=1}^N \frac{1}{|r_j-r|}
    \label{closeness}
\end{equation}
{$N$ is the total number of particles. The sensitivity factor $F_S$ in eq. \ref{fs}
depends on the average solid-to-gas ratio $\epsilon$ of the system:
\begin{equation}
F_s = \alpha (\epsilon - \epsilon_{\rm crit})  \quad \texttt{for} \quad \epsilon > \epsilon_{\rm crit}
\label{fs}
\end{equation} 
The solid-to-gas ratio $\epsilon$ is defined as the ratio between the total dust mass and the average gas mass,}
\begin{equation}
    \epsilon = \frac{N \cdot m_p}{V \cdot \rho_g} = \frac{1}{6} \pi s^3 \frac{N}{V} \frac{\rho_p}{\rho_g},
    \label{epsilon}
\end{equation}
{where $m_p$ is the mass of a single particle, $V$ is the total volume covered by particles, $\rho_g$ is the gas density in the chamber, and $\rho_p$ is the bulk density of the individual grains, and $s$ is the particle diameter.}

{As seen in eq. \ref{fs}, \citet{schneider2019} also empirically found that} particles {are only influenced by} the other particles in a cloud if the average  {solid-}to-gas ratio $\epsilon$ is above a threshold value $\epsilon_{\rm crit}${; otherwise, $F_s=0$ and particles sediment with $v_0$}. 

{The sensitivity $\alpha$ connecting the solid-to-gas ratio to the sensitivity factor in eq. \ref{fs} was just a constant in \citet{schneider2019}}.

{This description is purely empirical and was deduced from a single experiment so far.}
Here, we present a systematic analysis, where we varied the gas pressure, particle size, {and} rotation frequency of the chamber and improved the setup and data acquisition.

\section{Levitation Experiment}

\subsection{Setup}
\begin{figure}
    \centering
    \includegraphics[width=\columnwidth]{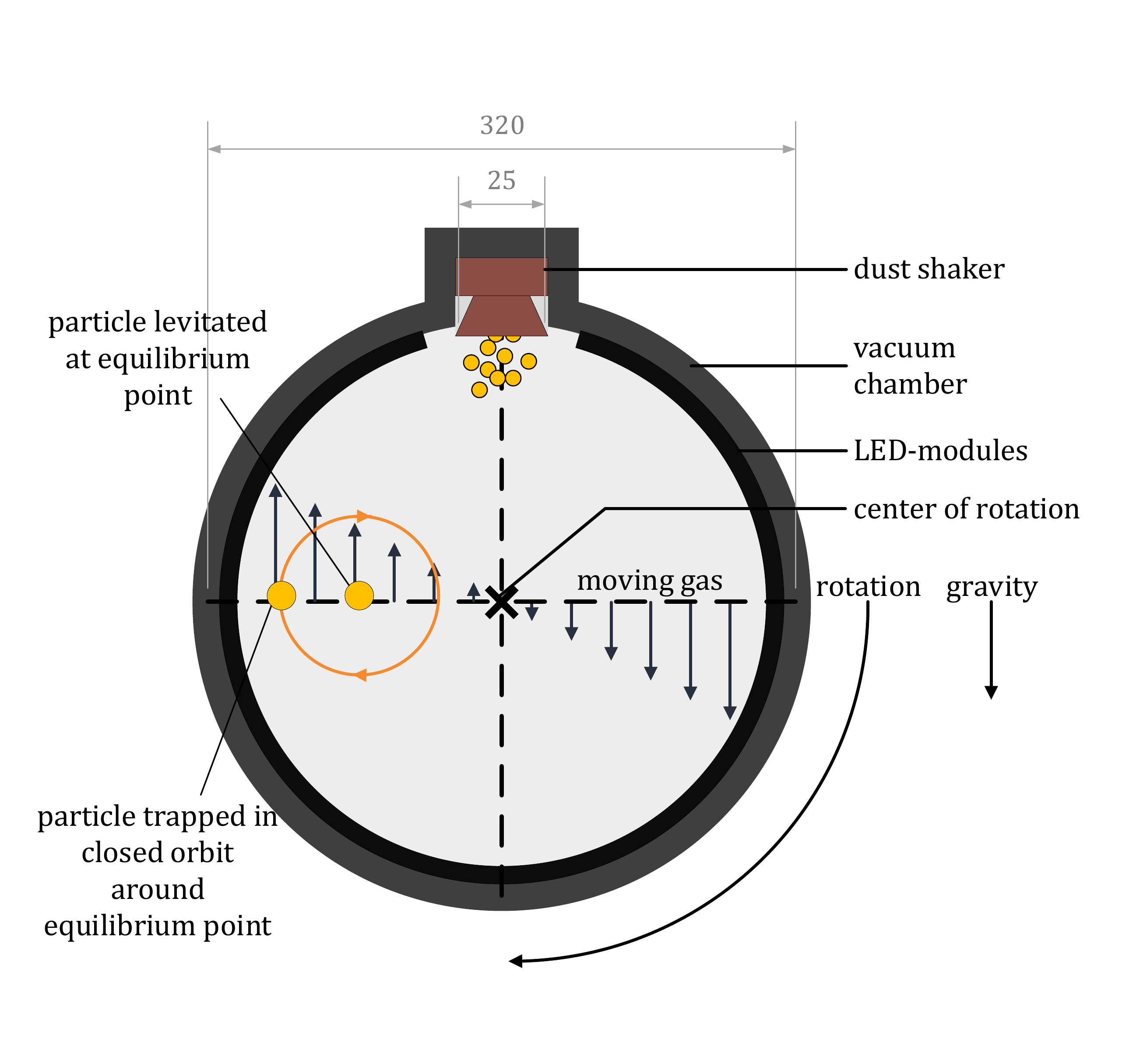}
    \caption{Experimental setup without auxiliary parts. The vacuum chamber is evacuated to a preset pressure. Two cameras observe the particles from the front. Illumination is provided by LED modules.}
    \label{fig:setup}
\end{figure}

The setup of the experiment (see fig. \ref{fig:setup}) follows the principle used in aggregation experiments by \cite{PoppeBlum1997} and \cite{blum1998}, but is especially based on earlier experiments on dense clouds by \citet{schneider2019}. 

Particles -- in this study, hollow glass spheres of different sizes and densities -- are dispersed {once at the beginning of the experiment,} within a rotating vacuum chamber with low ambient pressure. Particles are injected using a vibrating sieve in an extension of the vacuum chamber. The gas inside follows the rigid rotation of the chamber.
The vacuum chamber has a diameter of 320~mm. Inside the chamber, a ring of LEDs is used to illuminate the particles. The scattered light of the particles is detected by two non-rotating cameras.

{These} cameras observe the particles from the front at a distance of 40~cm on a 1'', 5~megapixel sensor with a spatial resolution in the order of 10 $ \rm \mu m $. The frame rate is 40 fps with an exposure time of 6 ms. The field of view is $19 \times 15 $cm. Imaging and synchronization of both cameras {are} controlled by a machine vision computer. Spatial calibration was realized with a calibration matrix with $\sim 10,000$  data points for each camera.

The experimental parameters are {the} particle radius of the sample $r_{\rm P}$, {the} particle bulk density $\rho_{\rm P} $, {the} gas pressure $p$,  and {the} rotation frequency $f$. We define the Stokes number of the experiment as
\begin{equation}
    \mathrm{St}=\tau_f \cdot f.
    \label{eq.St}
\end{equation}
The friction time is calculated with equation \ref{tau_f}.
The experiments were carried out similar to the ones described in \cite{schneider2019}. 
In short, the chamber was evacuated to a preset pressure and then disconnected from the vacuum pump. The injection process was then started and the chamber was set to a predefined rotation frequency before image acquisition for both cameras was started.

\subsection{Data analysis}

In principle, data were processed as in \cite{schneider2019}. We refer the reader to that paper for details. {All} particle positions were extracted for all times
with Trackmate \citep{trackmate}{,} using a Laplacian of Gaussian {(LoG)} particle detector and {a} Linear Motion LAP tracker for particle track assignment.
For data analysis, all particle positions and every particle track with track length $>100$ frames were taken into account.

{Since} we used two parallel cameras, the 3d position was reconstructed as {a} new feature here. The stereoscopic reconstruction was carried out by an algorithm that maps the expected particle position of the first on the second camera image and then finds matches by minimizing the difference between the projected position and detected particle positions of the second camera {image}.  

The error in the $z$-position of each particle is $\sim 2\% $.  
From these data, individual sedimentation velocities, individual closenesses, and average {solid-}to-gas ratios were determined. 

Furthermore, in this study, the sedimentation velocity was normalized to the undisturbed, individual sedimentation velocity of the grains used in the experiment $v_0$. The closeness was normalized by multiplication with the particle diameter $s$ of the glass beads used in the corresponding experiment (table \ref{tab:exp}).

\begin{figure}
    \centering
    \includegraphics[width=\columnwidth]{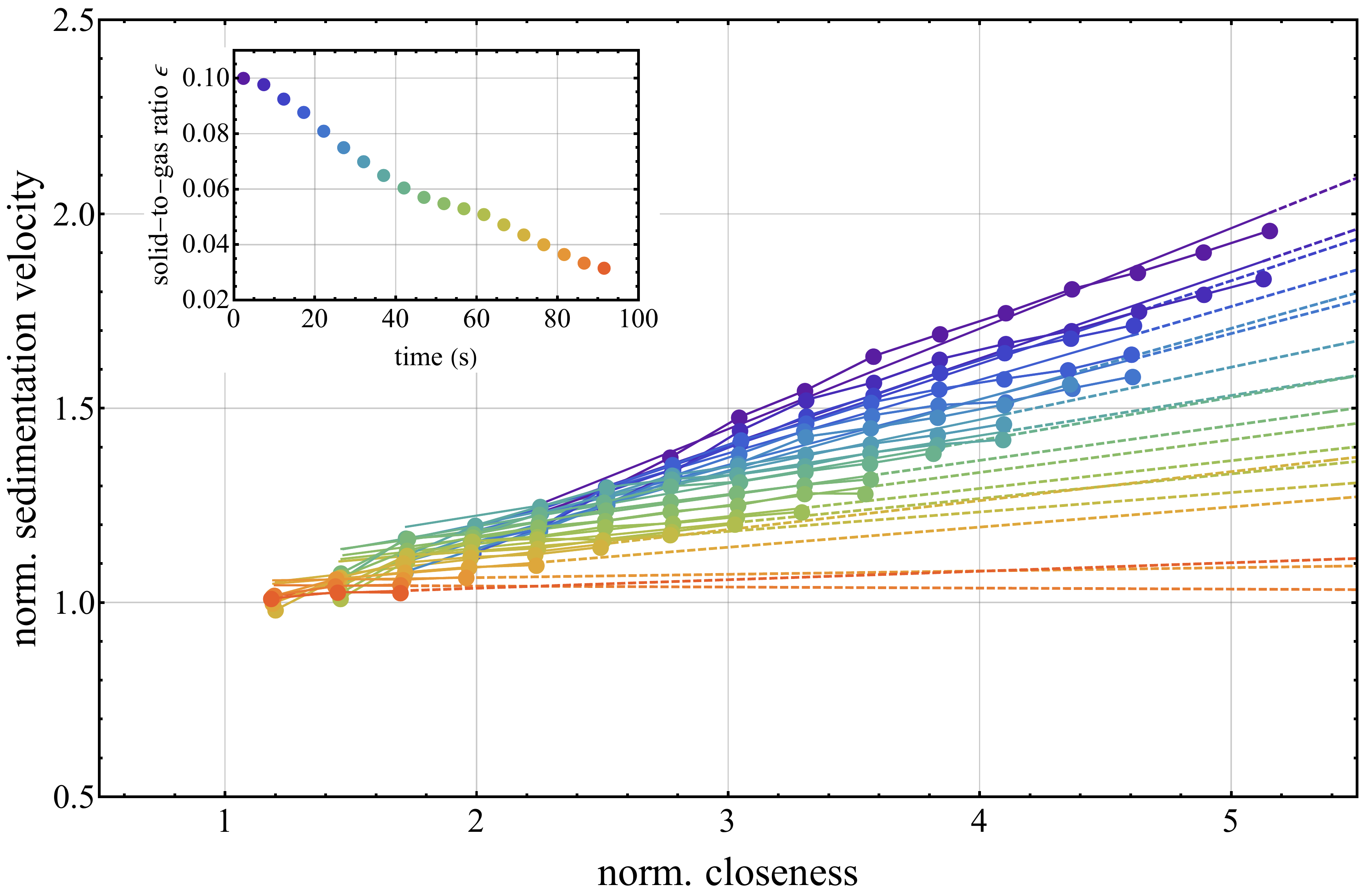}
    \caption{Normalized sedimentation velocity over normalized closeness for 19  {revolution}s of experiment 8 (table \ref{tab:exp}). The color of the data points refers to the  {revolution} of the experiment chamber during the measurement starting with  {revolution} 1 in blue (top) and  {revolution} 19 in red (bottom). Data points are average values for at least 1000 particle positions with an equidistant spacing of the binned values in closeness space. The total number of examined single sedimentation velocity data points is about 1,025,000; the total number of examined particle positions is about 1,600,000.\\
   {The top left inset shows the solid-to-gas ratio of each revolution as a function of time.}}
    \label{fig:sedimentationovercloseness}
\end{figure}

According to equation \ref{sedvel} and \ref{fs}, the sedimentation velocity depends on the closeness $C$ and the  {solid-}to-gas ratio $\epsilon$. Due to particle loss $\epsilon$ decreases with time. We {group} the measured particle positions and velocities in full revolutions of the experiment chamber. Fig. \ref{fig:sedimentationovercloseness} shows an example of the sedimentation velocity over closeness. This confirms the linear dependence found in \cite{schneider2019}.

The slope varies with every revolution or average $\epsilon$.
According to equation \ref{sedvel} and \ref{fs}{,} this slope is equal to $F_s = \alpha (\epsilon-\epsilon_{\rm crit})$.
 
\begin{figure}
    \centering
    \includegraphics[width=\columnwidth]{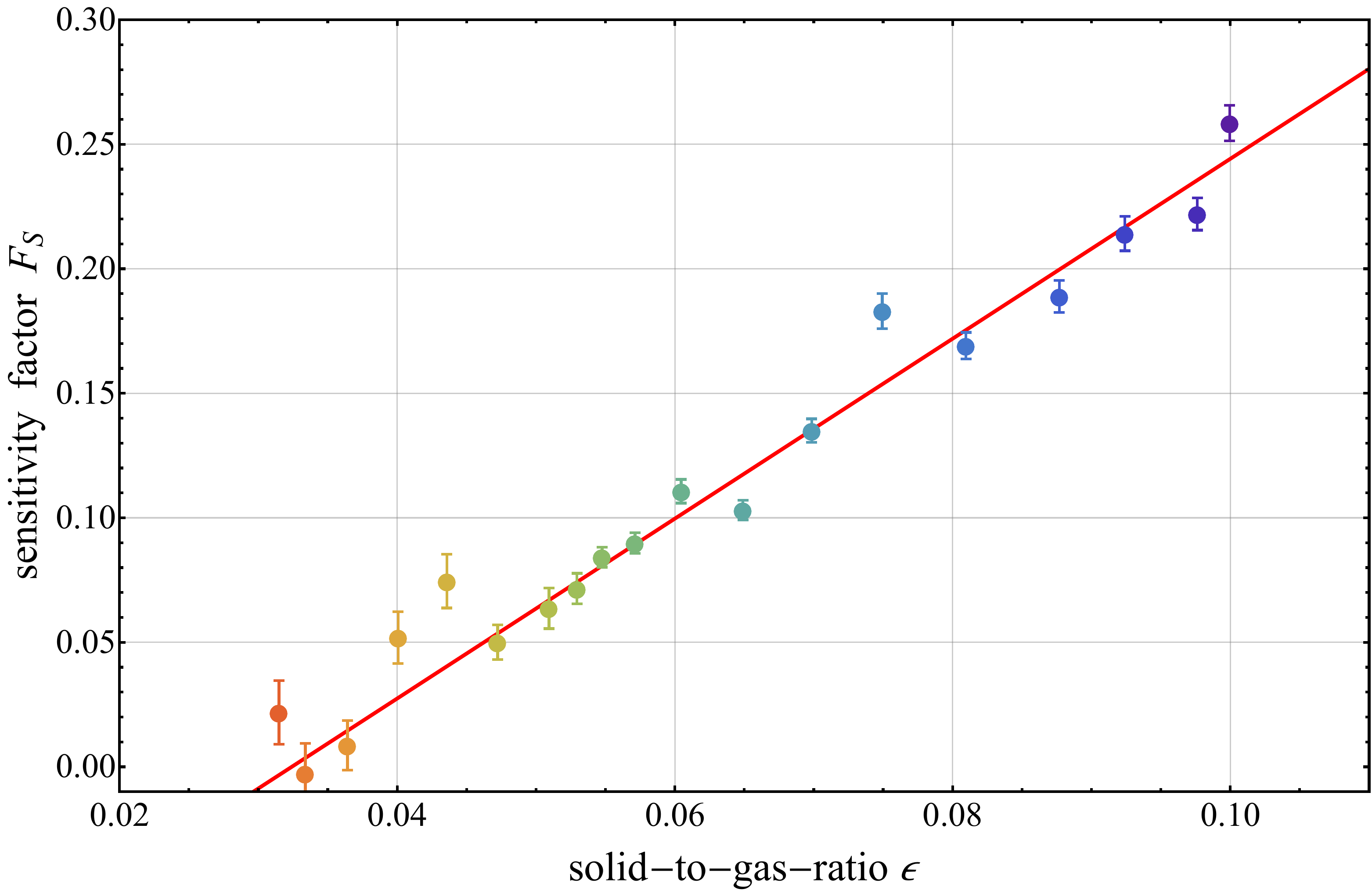}
    \caption{Sensitivity factor $F_S$ over solid-to-gas ratio $\epsilon$ of experiment 8 (table \ref{tab:exp}). The color of the data points refers to the data points shown in fig. \ref{fig:sedimentationovercloseness}. The linear fit is $F_S (\epsilon) = -0.12 + 3.6  \cdot \epsilon$}
    \label{fig:sensitivityoverepsilon}
\end{figure}

Fig. \ref{fig:sensitivityoverepsilon} confirms the linear trend of the sensitivity factor on $\epsilon$ \citep{schneider2019}. 
From the linear fit $F_S = a \cdot \epsilon + b$,   we can then deduce the sensitivity  $\alpha = a$ and the critical {solid-}to-gas ratio $\epsilon_{\rm crit}$  as
\begin{equation}
    \label{epscrit}
    \epsilon_{\rm crit} = -\frac{b}{a}.
\end{equation}

We define a system to be collective when individual sedimentation velocities deviate from $v_0$ or if $F_S >0$. We {define} a system as non-collective if all particles behave like test particles, sedimenting independently of local closeness variations.

\section{Discussion}

\begin{figure}
    \centering
    \includegraphics[width=\columnwidth]{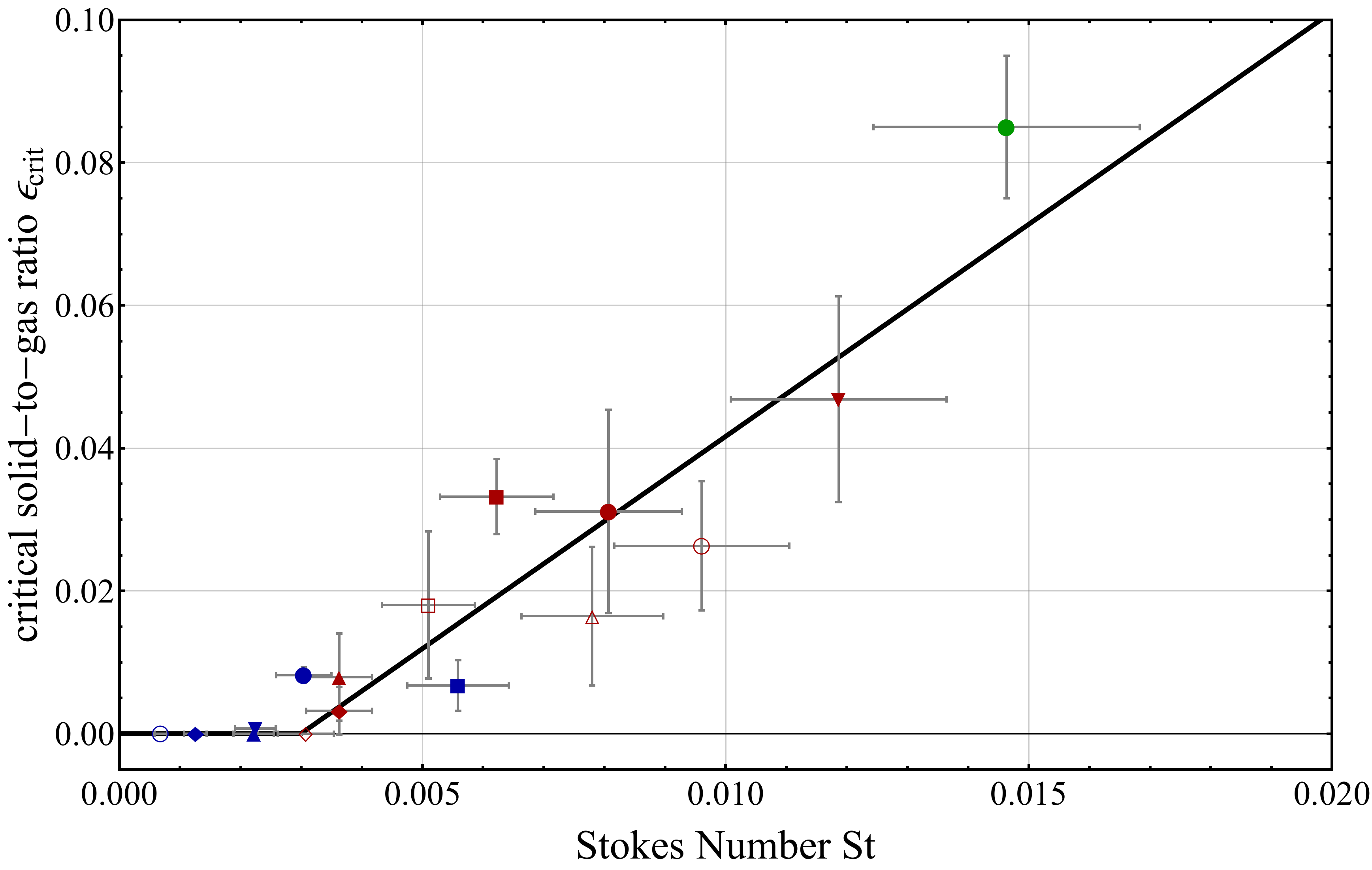}
    \caption{Critical solid-to-gas ratio of all performed experiments in dependence on the Stokes number. The shape and color of data points correspond to the experiments shown in tab. \ref{tab:exp}. Fit: $\epsilon_{\rm crit} (\mathrm{St}) = -0.018  + 5.9  \cdot \mathrm{St}$}
    \label{fig:epsilonoverstokes}
\end{figure}

After data analysis, two {main} quantities are given: $\epsilon_{\rm crit}$ and $\alpha$. The critical  {solid-}to-gas ratio varied for the different experiments carried out. As we also changed {several} parameters between individual experiments it is \textit{a priori} not  clear {whether} these two parameters follow systematic trends. {Therefore, we} considered $\epsilon_{\rm crit}$ to depend on a number of individual variables, including Knudsen number, pressure, and particle size. However, the only systematic dependence found was {concerning} the experiment's Stokes number $\rm St$, mainly influenced by $\tau_f$. This is shown in fig. \ref{fig:epsilonoverstokes}.

\begin{figure} [h]
    \centering
    \includegraphics[width=\columnwidth]{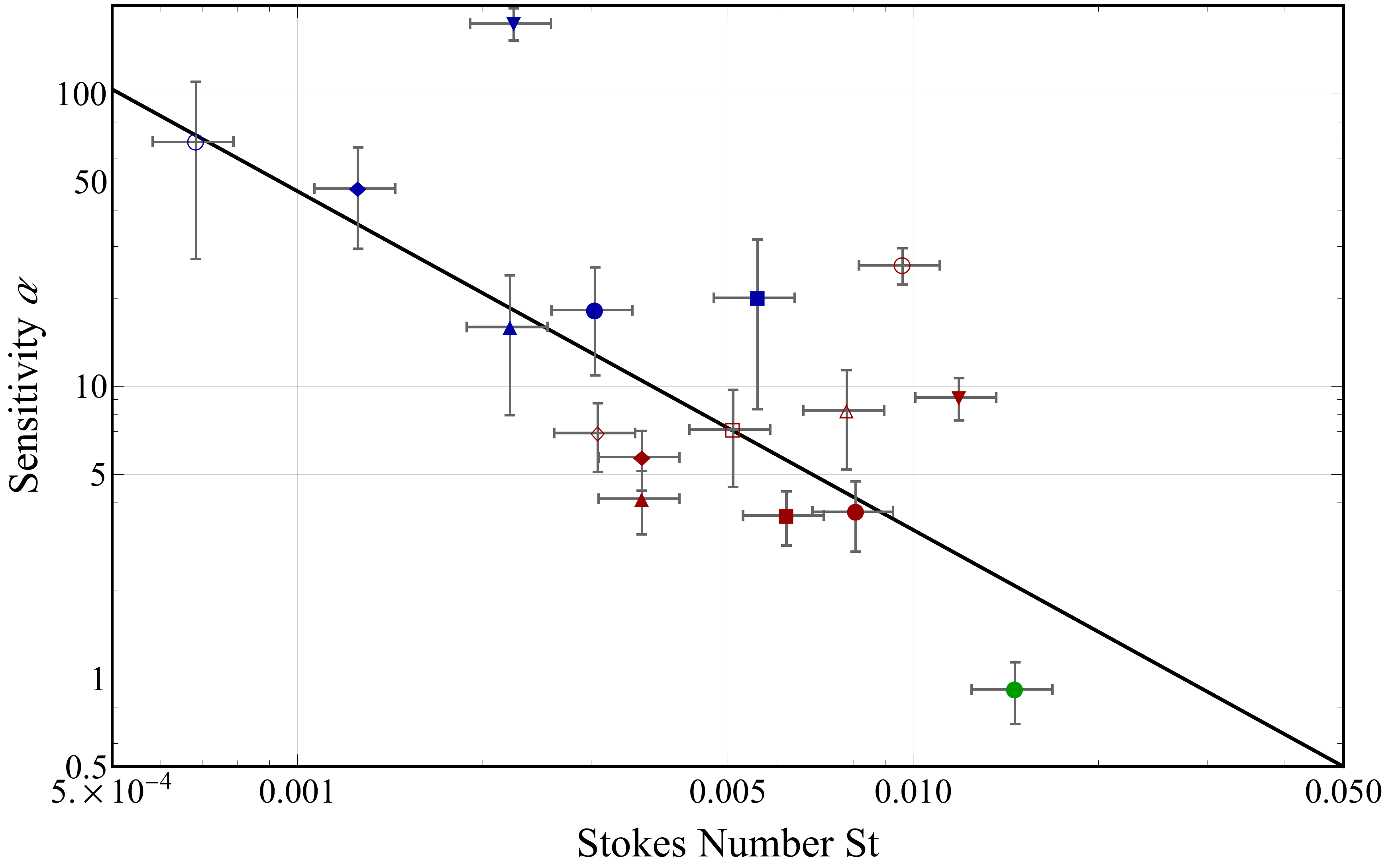}
    \caption{Sensitivity of all performed experiments in dependence on the Stokes number. The shape and color of data points correspond to the experiments shown in tab. \ref{tab:exp}. Fit: $\alpha (\mathrm{St}) = 0.016 \cdot \mathrm{St}^{1.2 }$}
    \label{fig:sensitivityfactoroverstokes}
\end{figure}

There is a clear linear trend in the data. 
{Interestingly, below a Stokes number of $\rm St \leq 0.003$, the deduced $\epsilon_{\rm crit}$ formally becomes negative. Negative $\epsilon_{\rm crit}$ refers to systems that are \textit{always} collective. Since $F_S$ always has to be larger or equal to 0, negative $\epsilon_{\rm crit}$ are set to 0}.
Particles with $\mathrm{St} \leq 0.003$ always back-react to the gas flow in such a manner that other particles are influenced by this.

The sensitivity $\alpha$
also depends on the Stokes number as shown in fig. \ref{fig:sensitivityfactoroverstokes}. We fitted a power law to the data as one possible functional dependence.
Small St particles have a higher impact on the gas flow than large St particles for the same average  {solid-}to-gas mass ratio.

The linear dependence of the sedimentation velocity on the closeness and on the solid-to-gas ratio is found for all parameter combinations. {Therefore, we consider this a robust, general finding.} 

\section{{Application to protoplanetary disks}}

{The Stokes number in protoplanetary disks is defined as $\mathrm{St} = \tau \cdot \Omega_{\rm K}$, where $\Omega_{\rm K}$ is the Kepler frequency. The Stokes number below which grains always behave collectively of 0.003 corresponds to particle sizes of about 1~cm at 1~AU for a particle density of 1~$\rm g cm^{-3}$ in a typical disk \citep{Johansen2014}.}

{For larger grains,} the system becomes increasingly insensitive to high  {solid-}to-gas ratios and only turns collective for higher values of $\epsilon$.
{It seems more than plausible that drag instabilities can only occur if the cloud becomes collective.
Therefore, this study suggests that grains {larger than 1~cm} require larger $\epsilon$ to trigger drag instabilities {at 1 AU or grains larger than 1~mm at 10~AU}.
For smaller grains{,} the clouds are always collective and very sensitive to changes in $\epsilon$. {Drag instabilities might therefore regularly occur for small grains rather than large grains. As grain growth proceeds in disks, pristine bodies might preferentially consist of entities of the threshold size, especially not of larger grains. This is in agreement {with} observations of comets \citep{Blum2017}.}}

\section{Conclusions}

Protoplanetary disks are regions with a wide range of solid--gas interactions, ranging from single test particle behavior of a dust grain in regions depleted of dust to solid dominated motion in {gravitationally} unstable particle subclouds. The  {solid-}to-gas mass ratio $\epsilon$ can vary from lower than interstellar $\epsilon \leq 0.01$ to larger than $\epsilon \geq 100$, while the volumetric filling factor $\Phi$ remains below $10^{-6}$. In our experiment, we confirm that the transition from test particle to collective behavior in comparably low-$\Phi$ environments is characterized by a threshold for the average  {solid-}to-gas ratio. Above the threshold, particle feedback on the gas is high enough to influence other particles.

This threshold depends on the Stokes number of the particles. The larger the Stokes number, the higher the  {solid-}to-gas ratio that still allows test particle behavior.
On the lower Stokes number end, our experiments \textit{always} come with collective behavior. Applied specifically to particle motion in protoplanetary disks, we would like to highlight two aspects of this work.  

First, in a simple cloud of small particles, their motion can be collective already at low  {solid-}to-gas ratios if the Stokes number is small, e.g. if grains are still dust and not yet pebbles. This, e.g., leads to increased sedimentation velocities. {As the maximum dust height is a balance between upward gas motion, e.g. as turbulent diffusion or convection, and sedimentation, faster settling corresponds to a reduced dust height for the same upward gas flow in parts behaving collectively. If this also changes the scale height observed astronomically depends on the local conditions at the respective height, i.e. if the top of the particle layer would be collective or non-collective. Collective sedimentation might also lead to a detachment of the surface layer and the midplane particles, but that is only a guess and further details are beyond the scope of this Letter.}  

{Also, other motions will change accordingly, e.g. the radial inward drift velocity for a given grain size in a collective ensemble will change}. 

Second, for large grains or rather at higher Stokes numbers, ever higher  {solid-}to-gas ratios are needed to get the cloud collective. 
{A threshold grain size of millimeter to centimeter marks the transition between always collective and solid-to-gas ratio dependence. Drag instabilities leading to planetesimal formation will {favor} this particle size supporting observations of comets. }

\section*{Acknowledgments}

This project is supported by DFG grant \mbox{WU 321/16-1}. We thank the two referees for a very constructive review of the paper. 

\newpage
\appendix

\section{Experimental Parameters}

\definecolor{dred}{rgb}{0.6,0,0}
\definecolor{dblue}{rgb}{0,0,0.65}
\definecolor{dgreen}{rgb}{0,0.6,0}

\begin{table}[H]
\begin{center}
\footnotesize {

\begin{tabular}{c|r|r|r|r|r|r|r|r|r|c}

\multicolumn{1}{c|}{Experiment}  & 
\multicolumn{1}{c|}{Particle} & 
\multicolumn{1}{c|}{Particle} & 
\multicolumn{1}{c|}{Pressure} & 
\multicolumn{1}{c|}{Rotation} & 
\multicolumn{1}{c|}{$v_0$} & 
\multicolumn{1}{c|}{Initial $\epsilon$} &
\multicolumn{1}{c|}{Kn} & 
\multicolumn{1}{c|}{St} &
\multicolumn{1}{c|}{Re} &
\multicolumn{1}{c}{Symbol} 
\\
\multicolumn{1}{c|}{} 
&\multicolumn{1}{c|}{Size} 
&\multicolumn{1}{c|}{Density} 
&\multicolumn{1}{c|}{} 
&\multicolumn{1}{c|}{Frequency} 
&\multicolumn{1}{c|}{} 
&\multicolumn{1}{c|}{} 
&\multicolumn{1}{c|}{} 
&\multicolumn{1}{c|}{} 
&\multicolumn{1}{c|}{} 
&\multicolumn{1}{c}{} 
\\
 & 
 \multicolumn{1}{c|}{($\mu \rm m$)} &  \multicolumn{1}{c|}{($\rm kg$  $\rm m^{-3}$)} & \multicolumn{1}{c|}{ (mbar) }&  \multicolumn{1}{c|}{($10^{-3} \times$ Hz)} &  \multicolumn{1}{c|}{($\rm mm s^{-1}$)} & 
 \multicolumn{1}{c|}{($10^{-3}$)} &
 \multicolumn{1}{c|}{} & 
 \multicolumn{1}{c|}{($10^{-3}$)}&
 \multicolumn{1}{c|}{($10^{-3}$)}&
 \multicolumn{1}{c}{} 

\\
\hline
\hline
Schn19+ & 165 $\pm$ 15      & 60 $\pm$ 6       & 9.5 $\pm$ 1     & 336 $\pm$ 2     & 68 $\pm$ 7  & 150 $\pm$ 20 & 0.08 $\pm$ 0.02   & 14 $\pm$ 2  &
7 $\pm$ 2&
\textcolor{dgreen}{$\medblackcircle$}
\\ 
1& 36 $\pm$ 9   & 280 $\pm$ 76       & 8 $\pm$ 1       & 216 $\pm$ 2     & 22  $\pm$ 2   & 42 $\pm$ 4 & 0.46 $\pm$ 0.16   & 3.0 $\pm$ 0.5  &
0.4 $\pm$ 0.2 &
\textcolor{dblue}{$\medblackcircle$}
\\
2& 36 $\pm$ 9   & 280 $\pm$ 76        & 8 $\pm$ 1       & 273 $\pm$ 2     & 32 $\pm$ 3   & 41  $\pm$ 4 &  0.46 $\pm$ 0.16   & 5.6 $\pm$ 0.8 &
0.6 $\pm$ 0.25&
\textcolor{dblue}{$\medblacksquare$}
\\
3& 36 $\pm$ 9   & 280 $\pm$ 76        & 13.5 $\pm$ 1    & 145 $\pm$ 2     & 14 $\pm$ 1 & 23  $\pm$ 2 & 0.30 $\pm$ 0.10   & 12 $\pm$ 0.2  &
0.4 $\pm$ 0.2 &
\textcolor{dblue}{$\medblackdiamond$}

\\
4& 36 $\pm$ 9   & 280 $\pm$ 76        & 14 $\pm$ 1      & 231 $\pm$ 2     & 15 $\pm$ 2  & 21  $\pm$ 2 & 0.26 $\pm$ 0.09   & 2.2 $\pm$ 0.3  &
0.5 $\pm$ 0.2 &
\textcolor{dblue}{$\medblacktriangleup$}
\\
5& 36 $\pm$ 9   & 280 $\pm$ 76        & 14 $\pm$ 1      & 173 $\pm$ 2     & 20 $\pm$ 2  & 10  $\pm$ 1 & 0.26 $\pm$ 0.09   & 2.2 $\pm$ 0.3  &
0.7 $\pm$ 0.3 &
\textcolor{dblue}{$\medblacktriangledown$}
\\
6& 36 $\pm$ 9   & 280 $\pm$ 76        & 10 $\pm$ 1      & 117 $\pm$ 2     & 9 $\pm$ 1   & 6.2  $\pm$ 0.6 & 0.37 $\pm$ 0.13   & 0.7 $\pm$ 0.1  &
0.2 $\pm$ 0.1 &
\textcolor{dblue}{$\medcircle$}
\\
7& 132.5 $\pm$ 8   & 75 $\pm$ 4     & 3.9 $\pm$ 0.4   & 293 $\pm$ 2     & 43 $\pm$ 4  & 145  $\pm$ 15 & 0.26 $\pm$ 0.04   & 8.0 $\pm$ 0.1  &
1.4 $\pm$ 0.4 &
\textcolor{dred}{$\medblackcircle$}
\\
8& 132.5 $\pm$ 8    & 75 $\pm$ 4    & 3.9 $\pm$ 0.4   & 203 $\pm$ 2     & 48 $\pm$ 5  & 107  $\pm$ 11 & 0.26 $\pm$ 0.04   & 6.2 $\pm$ 0.9  &
1.6 $\pm$ 0.4 &
\textcolor{dred}{$\medblacksquare$}
\\
9& 132.5 $\pm$ 8    & 75 $\pm$ 4    & 4 $\pm$ 0.4     & 153 $\pm$ 2     & 37 $\pm$ 4    & 56  $\pm$ 6 & 0.25 $\pm$ 0.04    & 3.6 $\pm$ 0.5  &
1.3 $\pm$ 0.3 &
\textcolor{dred}{$\medblackdiamond$}
\\
10& 132.5 $\pm$ 8    & 75 $\pm$ 4    & 4 $\pm$ 0.4     & 153 $\pm$ 2     & 37  $\pm$ 4   & 98  $\pm$ 10 &  0.25 $\pm$ 0.04    & 3.6 $\pm$ 0.5  &
1.3 $\pm$ 0.3 &
\textcolor{dred}{$\medblacktriangleup$}
\\
11& 132.5 $\pm$ 8    & 75 $\pm$ 4    & 8.1 $\pm$ 1     & 378 $\pm$ 2     & 49   $\pm$ 5  &68  $\pm$ 7 &  0.12 $\pm$ 0.02   & 28 $\pm$ 2  &
3.5 $\pm$ 0.9 &
\textcolor{dred}{$\medblacktriangledown$}
\\
12& 132.5 $\pm$ 8    & 75 $\pm$ 4    & 8.1 $\pm$ 1     & 375 $\pm$ 2     & 40   $\pm$ 5  & 51  $\pm$ 5 &  0.12 $\pm$ 0.02   & 9.6 $\pm$ 1  &
2.8 $\pm$ 0.7 &
\textcolor{dred}{$\medcircle$}
\\
13& 132.5 $\pm$ 8    & 75 $\pm$ 4    & 12 $\pm$ 1      & 199 $\pm$ 2     & 40   $\pm$ 4  & 48  $\pm$ 5 &  0.08 $\pm$ 0.01    & 5.1 $\pm$ 0.7  &
4 $\pm$ 1 &
\textcolor{dred}{$\medsquare$}
\\
14& 132.5 $\pm$ 8    & 75 $\pm$ 4    & 8.1 $\pm$ 1     & 160 $\pm$ 2     & 30  $\pm$ 3   & 36  $\pm$ 4 & 0.12 $\pm$ 0.02   & 3.1 $\pm$ 0.5  &
2.1 $\pm$ 0.5 &
\textcolor{dred}{$\meddiamond$}
\\

15& 132.5 $\pm$ 8    & 75 $\pm$ 4    & 8.1 $\pm$ 1     & 297 $\pm$ 2     & 41   $\pm$ 4  & 74  $\pm$ 7 & 0.12 $\pm$ 0.02   & 8 $\pm$ 1 &
2.9 $\pm$ 0.8 &
\textcolor{dred}{$\medtriangleup$}
\end{tabular}

   } 
\end{center}

\caption{Parameters of the Laboratory Experiments. Schn19+ gives the data published in \cite{schneider2019}}
\label{tab:exp}

\end{table}

\newpage

\bibliography{references}{}
\bibliographystyle{aasjournal}



\end{document}